\documentclass[aps,prb,reprint,twocolumn,superscriptaddress,floatfix,nofootinbib,longbibliography]{revtex4-1}

\usepackage{graphicx,amsmath,amsfonts,amssymb,amsthm,xr}
\usepackage{epsfig,amsmath,amssymb,color,dsfont,upgreek,physics}
\usepackage{mathrsfs}
\usepackage{mathtools}
\usepackage{bbold}
\usepackage{comment}
\usepackage{float}
\usepackage[bookmarks=true,colorlinks,linkcolor=OrangeRed,urlcolor=NavyBlue,citecolor=RoyalBlue]{hyperref}
\usepackage[dvipsnames]{xcolor}
\usepackage{physics}
\usepackage{amssymb}
\usepackage{amsmath}
\usepackage{amsfonts}
\usepackage{natbib}
\usepackage{hyperref}
\usepackage{graphicx}
\usepackage[dvipsnames]{xcolor}
\usepackage{bm}
\usepackage[normalem]{ulem}

\begin{document}

\title{Meandering conduction channels and the tunable nature of quantized charge transport}

\author{Benoit Dou\c{c}ot}
\affiliation{LPTHE, UMR 7589, CNRS and Sorbonne Universit\'e, 75252 Paris Cedex 05, France}

\author{Dima Kovrizhin}
\affiliation{LPTM, CY Cergy Paris Universite, UMR CNRS 8089, Pontoise 95032 Cergy-Pontoise Cedex, France}

\author{Roderich Moessner}
\affiliation{Max-Planck-Institut f\"ur Physik komplexer Systeme, 01187 Dresden, Germany}

\begin{abstract}
The discovery of the quantum Hall effect founded the field of topological condensed matter physics. Its amazingly accurate quantisation of the Hall conductance, now enshrined in quantum metrology, is topologically protected: it is stable against any reasonable perturbation. Conversely, topological protection thus implies a form of censorship, as it completely hides any local information from the observer. The spatial distribution of the current in the sample is such a piece of information, which however has now become accessible thanks to spectacular experimental advances. It is an old question whether an original, and intuitively compelling, picture  of the current flowing in a narrow channel along the sample edge is the physically correct one. Motivated by recent experiments \textit{locally} imaging the quantized current flow in a Chern insulating  (Bi, Sb)$_2$Te$_3$ heterostructure [Rosen et al., PRL 129, 246602 (2022); Ferguson et al., Nat. Mater. 22, 1100–1105 (2023)], we theoretically demonstrate the possibility of a broad `edge state' meandering away from the sample boundary deep into the sample bulk. Further, we show that varying experimental parameters permits continuously tuning between  narrow edge states and meandering channels all the way to incompressible bulk
transport. This accounts for various features observed in, and differing between, experiments. Overall, this  underscores the robustness of topological condensed matter physics, but it also unveils a  phenomenological richness hidden by topological censorship -- much of which we believe remains to be discovered.
\end{abstract}

\maketitle

\section*{Introduction}
A central pillar of topological physics is the bulk-boundary correspondence: topological properties of gapped bulk reflected in gapless state at the edge. In a picture due to Halperin \cite{Halperin_edge}, this explains the quantisation of the Hall conductance \cite{IQHE,IQHE_metrology,topobook}: a single chiral conducting edge channel has conductance $e^2/h$, regardless of its length or the geometry of the sample. Corrections to this could arise from backscattering into a counterpropagating channel. However, as such a channel is available only at the other edge of the sample, separated by the incompressible bulk, such backscattering is suppressed and quantisation is almost perfect.

This picture is  simple, beautiful -- and it accounts for the experimentally observed quantisation. However, it contains considerably more microscopic detail than is reflected in a single global transport coefficient. A long-standing experimental research effort by the von Klitzing group \cite{weis_IQHEcurrent}, as well as theoretical work \cite{Wexler_Thouless_1994,Tsemekhman_97}, has indicated that, in fact, the actual situation in quantum Hall effect (QHE) systems is different, with evidence for the current transport occurring far from the edge.

\begin{figure}
\centering
\includegraphics[width=\columnwidth]{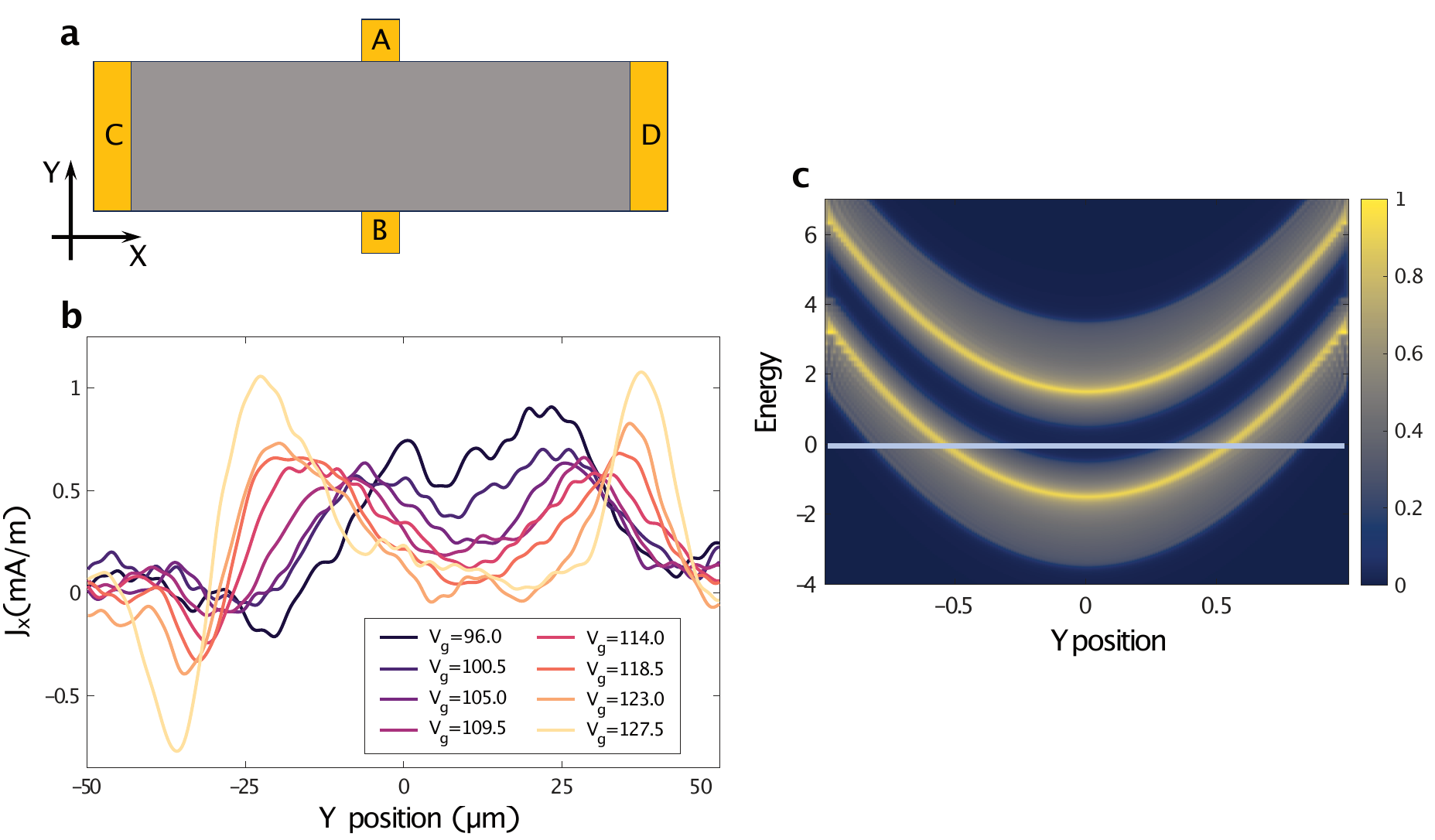}
	\caption{\textbf{Experimental motivation and central features of Chern band.}
 \textbf{(a)}:~schematic picture of the  quantum anomalous Hall effect (QAHE) experimental setup on the Chern insulator (Bi, Sb)$_2$Te$_3$ (red), where a transverse (Hall) potential $V_{AB}$ between points $A$ and $B$ is measured upon passing a current $I_{CD}$ between gold contacts $C$ and $D$. 
\textbf{(b)}:~experimental determination, Ref.~\cite{cornell_CIcurrent}, of the  distribution of the quantized horizontal current along a transverse cut through the sample for (Bi, Sb)$_2$Te$_3$: for different gate voltages, the current can be tuned smoothly between flowing near the edge and in the bulk of the sample.
\textbf{(c)}:~density of states of a disorder-free model for the Chern insulator hosting the QAHE in presence of a parabolic confining potential, with the y-axis  corresponding to the transverse direction (between points A and B in Fig.~\ref{fig:dichotom}a, with $V_p=5\times 10^{-4}$, $m=-1.5$, see Appendix). The chemical potential (horizontal line) corresponds to a locally empty (near the edge), incompressible (around the centre) and compressible region (in between). The meandering  channel is associated with the compressible region.%
}
 \label{fig:dichotom}
\end{figure}

Analogous quantized charge transport is also found in the  quantum anomalous Hall effect (QAHE) in Chern insulators \cite{Haldane_chern}, the experimental observation of which in (Bi, Sb)$_2$Te$_3$ \cite{QAHE_expt} has opened the field to microscopic investigation, Fig.~\ref{fig:dichotom}.  Indeed, an intriguing -- and technically highly impressive -- pair of contemporaneous  recent investigations on (Bi, Sb)$_2$Te$_3$ has also found a bulk contribution to the current, but of apparently strikingly different nature. Two samples \cite{cornell_CIcurrent,stanford_CIcurrent} suggested a rather  regular flow pattern essentially dictated by the solution of Laplace's equation in presence of a strongly anisotropic conductivity tensor; while another \cite{cornell_CIcurrent} found a richly structured flow pattern comprising broad channels reaching deep into the bulk. Depending on the gate potential, there was a single or a pair of channels, more or less separated by a central channel with lower current density.  

To account for these different observations, we develop a theory for the QAHE in Chern insulators which assigns a prominent role to the kinetic energy (non-vanishing bandwidth) of the Chern bands. This is in sharp contrast to the physics of the QHE, where the kinetic energy in a given Landau level is quenched, and therefore Coulomb interactions are relatively more important in determining the spatial density profile of the electronic system~\cite{Chklovskii_92,Wexler_Thouless_1994,weis_IQHEcurrent}. The two other ingredients in our model, also present in the QHE regime for 2D electronic systems are a smooth confining potential as well as a quenched random potential due to impurities -- with the latter yielding a quantized plateau in the QAHE via a different route than in the QHE.

So, while the complexity of the question where the current flows has been appreciated in the context of the QHE for a while, it is amusing to note -- for all it is worth -- that the artificial intelligence of ChatGPT gave us an unambiguous answer for the QAHE: ``In a Chern insulator [...] the current flows along the edges of the material, rather than through its bulk.''

\begin{figure}
\centering
\includegraphics[width=\columnwidth]{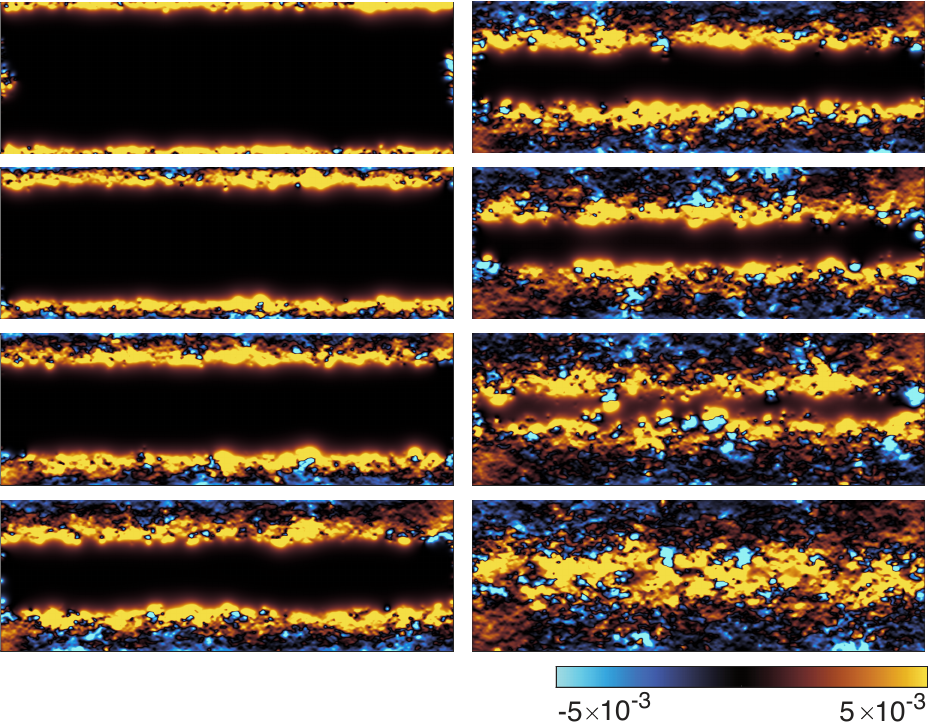}
\caption{
\textbf{Meandering channel and its continuously variable nature.} Plots show distribution of the horizontal component of the quantized transport current (excess current generated by application of the chemical potential difference) in Chern insulator in a Hall bar geometry. Data shown for eight chemical potentials ($\mu=0.6,0.2,-0.1,-0.3,-0.4,-0.5,-0.6,-0.7$, from top left to bottom right); parabolic potential $V_p=5\times 10^{-6}$ and disorder strength $V_d=1.25$ (system size $=1000\times 3000$ unit cells). The results presented are shown after applying a Lorentzian filter which mimics the spatial resolution of the experimental measurements. Corresponding charge density distributions are shown in Appendix Fig.~\ref{fig:density_panels}.}
	\label{fig:currentpattern1}
\end{figure}

We find, visually strikingly (Fig.~\ref{fig:currentpattern1}), and at variance with a simple picture of edge state transport, what we call a {\it meandering conduction channel}, which can both be broad and spread deep into the bulk.  Its width, which is typically much larger than any microscopic length, is set by the interplay of the band kinetic energy and the confining potential, with a smaller influence of disorder strength.

Varying experimental parameters thus allows one to access a number of different regimes {\it all} exhibiting quantized transport.  We can in particular explicitly account for the current distribution of Ref.~\cite{cornell_CIcurrent}, as shown in Fig.~\ref{fig:compr_incompr}. Our picture also resolves several  questions thrown up by experiment and (apparent) theoretical dichotomies. Needless to say, the topological stability of the QAHE ensures that all of these yield the correct quantized transport coefficient, even if their detailed realisations are hugely different. 

\section*{Results}

We have obtained spatial current distributions in a Chern insulator in a two-terminal geometry from our analysis of a standard model for the QAHE \cite{Qi_QAHE}, in the presence of disorder and a parabolic confining potential (see Appendix Fig.~\ref{fig:parabola}). 

We can readily identify several strikingly different regimes in Fig.~\ref{fig:currentpattern1}. First, there is the  case of a `conventional' narrow  channel at  each edge. These channels turn out to be continuously tunable by simply sweeping the chemical potential (accessible experimentally via backgate voltage): they move into the sample, and broaden in the process. Eventually they meet in the sample centre without any apparent spatial separation, at which point  edge of the quantisation plateau is reached, as the quality of the quantisation starts to degrade notably.

The existence of these different regimes within a single theoretical treatment is the first central result of this work, as it accounts for the large microscopic variability of which the quantized transport is independent, and which -- in the absence of local probes -- is  masked by topological censorship.

\subsection*{Meandering conduction channel}

Our second central result is implied by the first: the completely natural appearance of a broad conduction channel in the QAHE. 
It  arises as the combination of the Chern bands of finite width separated by a gap on one hand, and the confining potential on the other, yield a spatially inhomogeneous set-up in the following way.
Locally, the chemical potential can lie below both bands; within the lower (valence) band; between the bands; within the upper (conduction) band; or above both bands (Fig.~\ref{fig:dichotom}). Most basically, the conduction properties of each situation are different, and hence changing the chemical potential will also change where the current flows. 

Consider the horizontal line in {Fig.~\ref{fig:dichotom}},
where we sketch the local density of states of a clean system as a function of the transverse spatial coordinate. {Here, due to the potential, the energy at the bottom of the valence band near the edge lies in the gap between valence and conduction band in the sample centre.} Placing the chemical potential at this energy, there
must then be two spatially extended regions either side of the centre, where the chemical potential lies within the valence band. Quantum mechanically, these regions host equal and opposite equilibrium anomalous  Hall currents \cite{Self_current}, turning these into conducting channels with {\it intrinsically finite widths}.  Near the boundary, no current carriers are available as the confining potential pushes the bottom band above the chemical potential. 
At the same time, the middle region, which is incompressible when  the chemical potential lies between the bands, can also carry current in presence of Coulomb interactions, as we  discuss below. 

We  turn to the non-equilibrium current, i.e.~the excess current distribution due to a chemical potential difference $\Delta$ between the lateral ends of the sample (denoted C, D in Fig.~\ref{fig:dichotom}a). 
This leads to redistribution of charge in the two meandering channels, see Appendix. However, while in QHE systems we have purely chiral edge states, this is different for the QAHE, where each Chern band contains states propagating in both directions (which we refer to as propagating and counter-propagating) -- rendering current quantisation more delicate. 
{How, then,  is quantisation achieved for the QAHE?}

\subsection*{Role of disorder}

To stabilise the quantized conductance it turns out that -- as in the QHE -- disorder plays a crucial role, see Appendix Fig.~\ref{fig:quantisation}.
In a nutshell, disorder Anderson-localises all states which are not protected by their chiral nature. This in particular removes the capacity of all counterpropagating states to carry current. Crucially, firstly, the number of states which survive localisation is given by the Chern number; and, secondly, their geometrical appearance/spatial distribution can vary wildly depending on disorder. Indeed, the non-equilibrium current distribution depicted in Fig.~\ref{fig:currentpattern1} (before coarse-graining) has microscopic fluctuations greatly in excess of the mean current densities. For the more detail-oriented reader, we emphasize with a more expansive discussion in Appenddix how differently  disorder asserts itself in the QAHE in comparison with the QHE.

Generically, the current-carrying region can have the same support as the region where the chemical potential resides in the Chern band: {\it the appearance of a broad current-carrying region is entirely natural in this picture.} Its width is eventually determined by disorder, and primarily by a combination of bandwidth of the Chern band, and steepness of the effective potential in the region -- itself externally tunable -- where the chemical potential intercepts the Chern band. Our results for the four-terminal resistance  are consistent with a previous analysis of the two-terminal situation in a disordered Chern insulator in a transverse electric field~\cite{QAHE_steppotential}.

\subsection*{Role of Coulomb interactions}

The full disordered and interacting problem, as usual, does not allow for an analysis on the scale and detail of our above treatment of non-interacting disordered electrons. We can, nonetheless, make a number of robust qualitative observations, which turn out to be conceptually important. 

Without the Coulomb interactions, the quantized non-equilibrium current is carried entirely by the additional electrons and holes in the respective compressible strips corresponding to meandering channels. The effect of adding Coulomb interactions is then (at least) twofold -- we analyse a simple illustrative phenomenological model  in Appendix, and here mention the results of importance for the following discussion of experiment.

\begin{figure}
\centering
\includegraphics[width=\columnwidth]{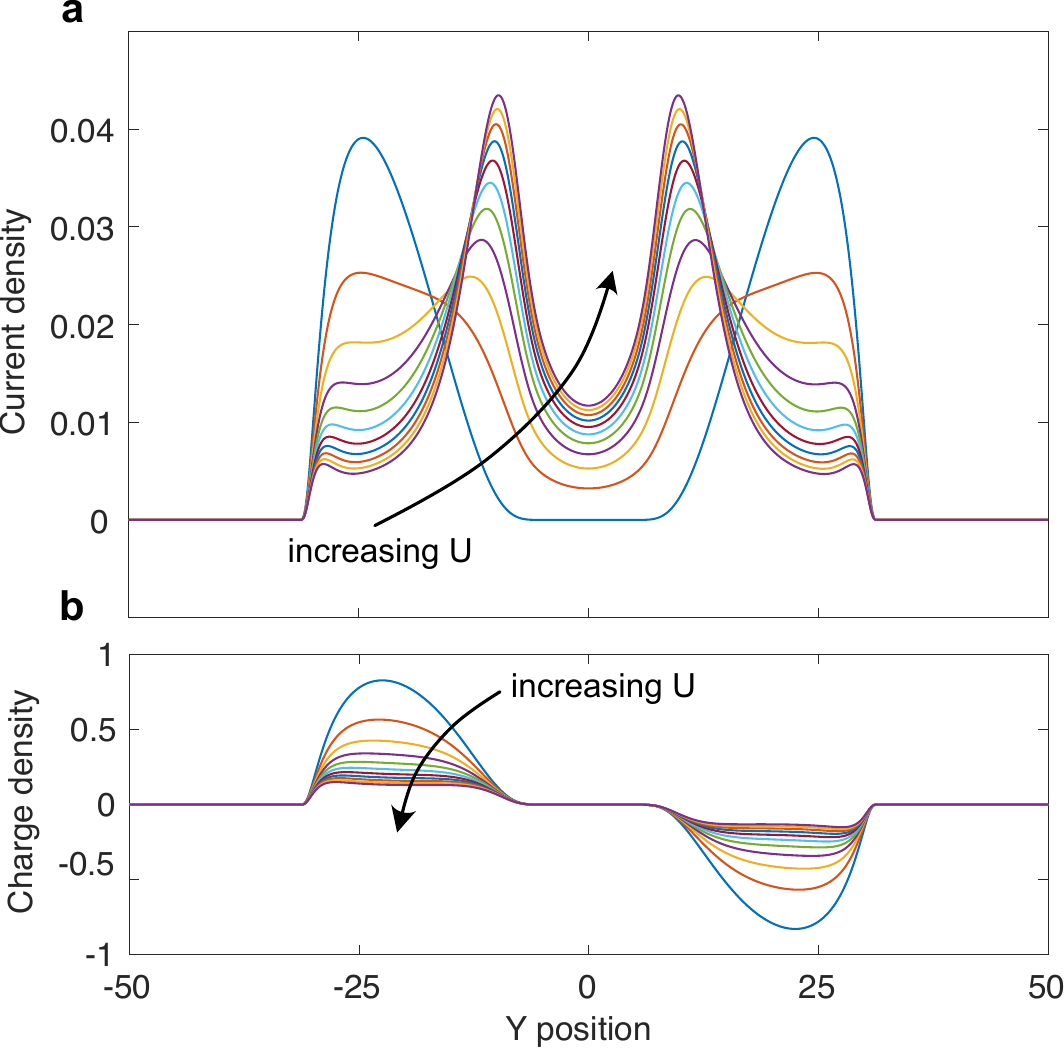}
\caption{
\textbf{Influence of Coulomb interaction:} change in \textbf{(a)} current  and \textbf{(b)} charge  density: our simple phenomenological model  shows transport current moving inwards in the compressible region, with the simultaneous appearance of transport current in the incompressible bulk  around $Y=0$ (where the charge density is unchanged), as the  interaction strength  increases from $U=0$ (blue line) to $U=0.002$ in equal steps.}\label{fig:currentpatternCoulomb}
\end{figure}

First, within each channel (Fig.~\ref{fig:currentpatternCoulomb}), the charge distribution flattens, while the peak of the current distribution shifts towards the center of the sample, with the inner edge carrying an increasing amount of current on account of the higher equilibrium density of electrons there. The second is qualitatively even more striking: the added electrons and holes in the two respective channels generate electric field in the incompressible region separating them, just like a charged capacitor consisting of two parallel plates; thus inducing a quantized bulk current there, which is absent in the non-interacting model ($U=0$ versus $U>0$ curves in Fig.~\ref{fig:currentpatternCoulomb}). (We find that the current in the central incompressible region is strongly suppressed by the screening due to a nearby metallic gate.) With this in hand, we can discuss the following central question.

\subsection*{Determining the  current distribution, and tuning between different regimes}

The  question regarding the qualitative nature of current flow is often (but not always \cite{Wexler_Thouless_1994,GirvinLesHouches1998}) phrased in the form of a dichotomy: \textit{does the current flow in the incompressible bulk or along the compressible edge?} Our answer to this question is twofold, and it negates the strict dichotomy: firstly, it depends on the detailed circumstances; and secondly, the current can do both at the same time. The only thing that matters for quantisation is the total electrochemical potential drop, whose detailed shape is fixed  by electrostatics, disorder and the Coulomb interaction. Only the  distribution of the current is influenced in this way, not its total value. This is  topological censorship at work.

\begin{figure}
\centering
\includegraphics[width=\columnwidth]{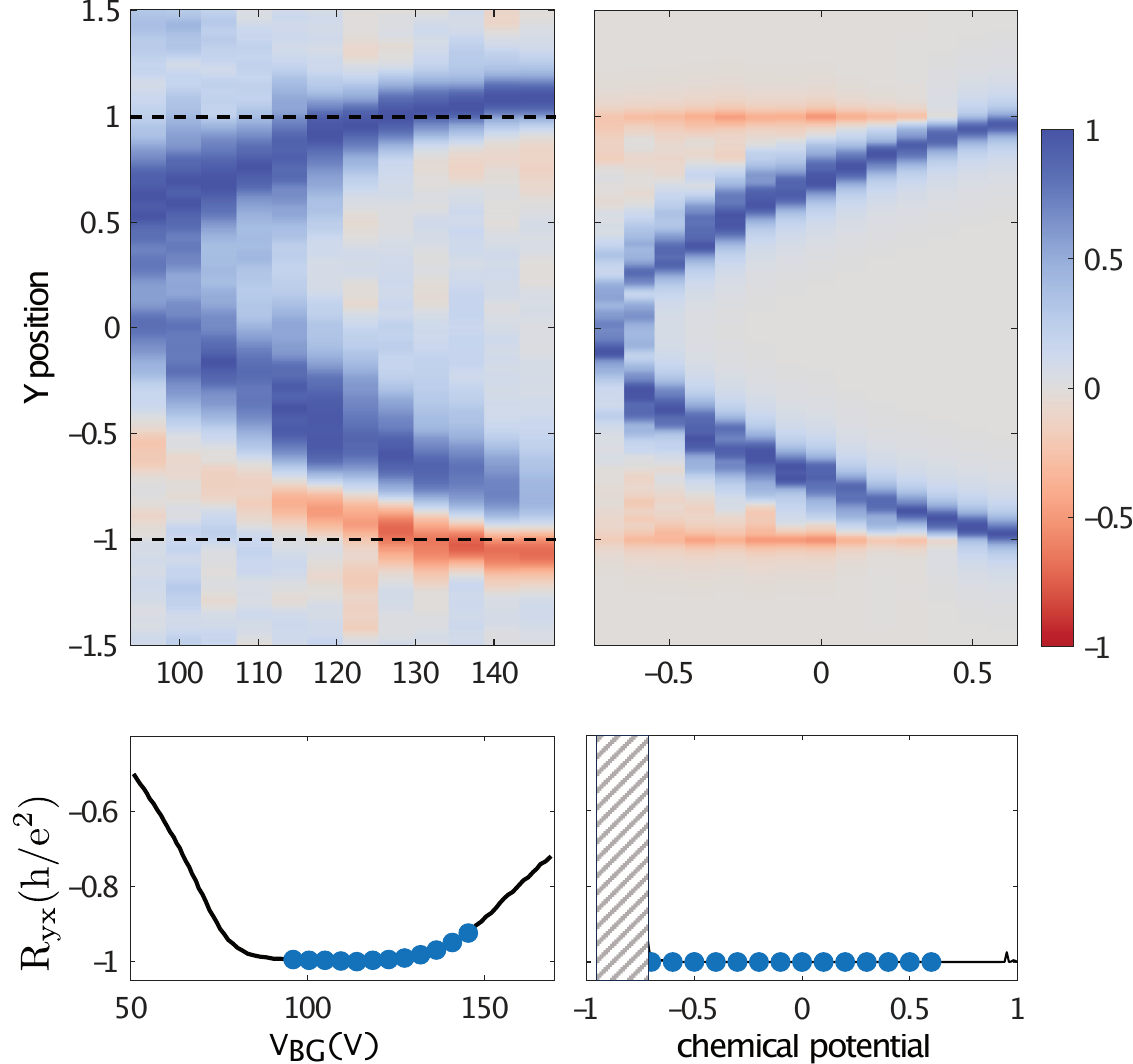}
	\caption{\textbf{Experiment-theory comparison of the spatial current density distribution in the QAHE.} Left: experiment, Ref.~\cite{cornell_CIcurrent}, right: our theory, as in Fig.~\ref{fig:currentpattern1}. Top panels show a false colour plot of the normalised current across the sample, vertical axis, against  the range of backgate voltages (experiment), or chemical potential (our theory). We only show data where the current is well-quantized, as indicated by the blue points in the bottom panels; quantisation breaks down chaotically in the shaded area. At each voltage/chemical potential, we normalise the density by the maximum of the absolute value. The vertical axis is rescaled by the half-width of the system, so that the physical sample boundaries correspond to $\pm1$, indicated with dashed lines. For further details of comparison, see Appendix.}
	\label{fig:compr_incompr}
\end{figure}

\subsection*{Analysis of experiment} We are now in a position to connect our theoretical picture with  beautiful experimental results of Refs.~\cite{stanford_CIcurrent,cornell_CIcurrent} on the QAHE in Cr-doped (Bi,Sb)$_2$Te$_3$ heterostructures. The Stanford experiment~\cite{stanford_CIcurrent} measured the potential along the edge using local (lateral) voltage probes. It found the potential profile consistent with the current distribution arising from a translationally invariant and strongly anisotropic conductance tensor, of the type discussed in Ref.~\cite{Gerhardts_03}. This experiment seems to rule out the picture of the current flowing exclusively through a narrow edge channel, but it says little about the microscopic nature of the bulk and the conduction process through it, and indeed about the origin and the scales of the nonzero diagonal entries of the conductance tensor. As usual, diagonal components of this tensor are associated with dissipation processes.

The Cornell experiment~\cite{cornell_CIcurrent} used a micro-SQUID to measure the magnetic field distribution above the sample. From this, the current distribution was inferred for different experimental parameters, which is depicted in Figs.~\ref{fig:dichotom}b~and~\ref{fig:compr_incompr}. The theory-experiment agreement in the latter is our final central result.

The most striking result of these measurements is the observation of a pair of broad regions in the bulk -- rather reminiscent of our meandering channels -- carrying the non-equilibrium current in the quantized regime. This is again at variance with a simple picture of charge transport in a narrow edge channel, while it also suggests considerably more spatial structure than the Stanford experiment. Given details of the sample geometry, preparation, gating different between these experiments, our observation of the non-universality of the transport mechanism is plausibly supported by this discrepancy.   

The authors of Ref.~\cite{cornell_CIcurrent} suggest an interpretation in which this current is carried entirely in the incompressible regions. They support this claim by emphasizing that regions carrying most of the non-equilibrium current are also those in which the magnetic response to a modulation of the gate voltage reaches its maximum absolute value. The sequence of current profiles in the quantized regime  is therefore 
interpreted as follows. The curve corresponding to the highest chemical potential on the quantized Hall plateau has two peaks, attributed to two incompressible strips. These strips gradually move towards the center of the sample as the chemical potential is reduced, and finally merge.

Based on our discussion of meandering channels, we do not see any particular reason why there should be no transport at all in the compressible regions, and at any rate, the experimental data is not precise enough to exclude this possibility. In fact, the current never seems to
entirely vanish near the center of the sample, which, according to the proposed picture, is supposed to host a partially filled (hence compressible) conduction band in a finite range of chemical potentials. Also, our attempts to derive a mechanism of a purely incompressible bulk transport for realistic models, based on a Hartree-Fock treatment, were unsuccessful. For an explicit illustration of the possibility of simultaneous transport in compressible and incompressible regions we refer to the above-mentioned (Fig.~\ref{fig:currentpatternCoulomb})  phenomenological model in Appendix.

Our results on the quantized Hall response carried by compressible meandering channels then suggest an alternative interpretation.
The large peaks are now attributed to such compressible meandering channels built from states lying inside the valence band. The residual current near the center might result from the transverse electric field across an incompressible region, due to charge imbalance between the two meandering channels, whose small value is compatible with the strong screening of Coulomb interactions due to the close proximity ($\sim$ 40~nm) of the conducting top gate. When the chemical potential is reduced, the meandering channels move gradually towards the center. As they do so, they encounter a weakening local electric field due to the confinement potential, and hence broaden.

In passing, we note the asymmetry of the experimental non-equilibrium current, showing a sign-changing structure on one side, and a uniform sign on the other. This appears to  involve heating, i.e.\ physics beyond the linear response regime \cite{cornell_CIcurrent}. 
A sign-changing pattern is also promoted by the fact that the Berry curvature of each  Chern band changes sign as a function of energy, see Appendix {Fig. \ref{fig:berry_curvature}}. In presence of confining potential and disorder, this translates into a tendency to carry current in opposite directions on opposite sides of the meandering channel; this appears to explain the red traces near the sample edge in the right panel of Fig.~\ref{fig:compr_incompr}.

\section*{Discussion} Confronting the new microscopic experiments on the QAHE with the lore of theory of quantized Hall conductance has in our minds  sharpened the following considerations. 
To start with, B\"uttiker's theory of transport formulated in terms of conducting edge channels \cite{Buttiker_edge} does not {\it per se} require the existence of states which have the visual appearance of a narrow edge channel. Rather, a broad and meandering  channel, resembling  a stream flowing in its marshy floodplain rather than in a sterile canal, is accommodated perfectly satisfactorily. This then  removes the notion that a channel meandering far from the edge into the bulk necessarily implies physics beyond B\"uttiker. It does, however, not mean that ingredients such as interactions -- a priori absent from the B\"uttiker formalism -- are unimportant. But their actual role may again vary considerably in different settings. 
The main differences appear in the nature and physical properties of {\it compressible regions.} In the QHE regime,
the electronic kinetic energy in a given Landau level is the same for all states. As a striking consequence, the spatial electronic distribution displays an alternation between compressible and incompressible stripes~\cite{Chklovskii_92}. For the compressible regions to be energetically stable, the local electrostatic potential has to
adjust self-consistently to remain constant across them. Therefore, it is generally concluded that they carry no Hall 
current~\cite{weis_IQHEcurrent,Tsemekhman_97,Gerhardts_03}. By contrast, the Landau level filling factor is constant across
incompressible stripes, and the electrostatic potential drop across them is responsible for almost all of the quantized Hall 
response, the small remaining part being carried by sharp edge states (on the scale of the magnetic length)~\cite{weis_IQHEcurrent,Tsemekhman_97,Gerhardts_03}.

In the QAHE, compressible regions created in the presence of smooth
confining potentials behave very differently, in that the local electrostatic potential may now vary across them.
This possibility may at first sight seem to imply absence of current quantisation but as we demonstrated above, the current of the compressible region itself can also be quantized. We believe that this insight compels us to  reconsider the status of such compressible regions in the QAHE. For instance, it is neither necessary for such regions to be absent \cite{stanford_CIcurrent}, nor does quantisation preclude an electrostatic potential gradient across them \cite{cornell_CIcurrent}. 

Further in this vein, it is amusing to note that two popular, physically different mechanisms of charge transport, can coexist in different parts of the sample. These are, first, population imbalances due to chemical potential differences, as would  occur even for charge-neutral systems such as cold atoms, and which are sufficient for narrow edge channel transport. And second, driving charged particles via forces ultimately due to an applied electric field, which would be all that is needed to account for the quantized current carried by the incompressible bulk. 

Note the crucial role played by disorder also for the QAHE. It is needed to localise the counterpropagating modes in the Chern bands, which would otherwise spoil current quantisation in the compressible regions. As in the QHE, but via a different mechanism, simple disorder-free models do appear to provide the correct quantized current, but do not produce a robust plateau.  

While technologically challenging, a way of probing the relative relation of current flow and potential distribution  would be to carry out simultaneous measurements of the potential and current distribution, i.e.\ by merging the experimental approaches of Ref.~\cite{weis_IQHEcurrent} and Ref.~\cite{cornell_CIcurrent}.  
A sharp question would be whether current flow exists in a region without appreciable electrochemical potential drop, as would be appropriate for the meandering channel. It may also be worth noting that it will be interesting to study how local fluctuations increase with  improved spatial resolution, possibly yielding insights into properties of the disorder potential.

\section*{Outlook}
We have shown how a topologically stable `universal' phenomenon of fundamental historical and conceptual importance can microscopically be realised in many, very  different, `non-universal' ways. This is a testament to the stability of topology to deformations which may be physically striking even if they do not change the topologically fixed observables, a point perhaps somewhat obscured by the striking beauty \cite{Wen_IQHEedge} of more abstract treatments of the subject.
Our results also reaffirm the crucial role played by disorder for the  stability of the quantisation \cite{QAHE_steppotential}, and its independence of non-universal features of the system. 

More broadly, we believe that our work suggests that a more systematic search for different phenomenologies underpinning the same topological response is severely underexplored and likely holds many surprises. 
This is a most timely endeavour since  recent rapid progress in the development of local probes \cite{Zeldov_QHEequil,Zeldov_nonlocal,Yacoby_graphenehydro,Ilani_hydro} can now be leveraged to attack the previously inaccessible local aspects of topological systems -- which may involve the appreciation of the physics taking place on various local scales. Their understanding can in turn feed into a higher degree of optimisability and controllability of designed nanosystems, such as the moir\'e systems, which are seeing tremendous activity \cite{Sid_FQAHE_2024} with the recent observation of the fractional QAHE \cite{FQAHE_park,FQAHE_zeng}.

\section*{Appendix}

The numerical results presented in the paper were obtained using a standard theoretical model of a Chern insulator introduced in Ref.~\cite{Qi_QAHE}. We consider a single-particle Hamiltonian on a square lattice with dimensions $N_x,N_y$, in presence of a parabolic confining potential and on-site disorder. The Hamiltonian describing a Chern insulator reads,
\begin{multline}
\hat{H}_0=\sum_{n}\hat c^{\dagger}_{n}\frac{\hat\sigma_z-i\hat\sigma_x}{2}\hat {c}_{n+\hat x}+\hat c^{\dagger}_{n}\frac{\hat\sigma_z-i\hat\sigma_y}{2}\hat {c}_{n+\hat y} +\mathrm{h.c.}\\+m\sum_{n}\hat{c}^{\dagger}_{n}\hat{\sigma}_z\hat{c}_{n}\ .
\end{multline}
Here, $\hat{c},\hat{c}^{\dagger}$ is a fermionic annihilation (creation) operator, and $\hat{\sigma}_\alpha$ a set of Pauli matrices. 
This model has been extensively studied in the case of clean systems in equilibrium \cite{Prodan_2009}, and it provides a good qualitative description of the BST Chern insulators studied in the above experiment. Throughout the calculations we set $m=-1.5$. 

For the calculations of non-equilibrium behaviour, such as the four-terminal conductance and current profiles, we use the KWANT package \cite{KWANT}. All of the results are presented for lattices of dimension $N_x=3000,\ N_y=1000$, parabolic potential strength $V_p=5\times 10^{-6}$, and i.i.d. on-site disorder $V_d$ with a box distribution on   the interval $0...2$. The results for the non-equilibrium current and the density were obtained for a single random disorder configuration by taking a sum over $100$ states in the energy window of $\Delta=0.02$ around a given energy. For the system size  used, there are about $1000$ scattering states at a given energy, so that the current is obtained from a sum containing ~$~10^5$~contributions. In order to calculate the current, we apply a chemical potential difference on the two sides of the system by occupying/deoccupying scattering states on the left/right side of the sample within the energy window $\pm \Delta/2$, respectively, around the equilibrium chemical potential $\mu$.

To make contact with experiment, the raw data for the current were post-processed via a convolution with the Lorentzian function,
\begin{equation}
F(r)=F_0(a_0+a_1/(r^2+a_1))e^{-\alpha r},
\end{equation}
where the parameters $a_0=5,\ a_1=165$ were obtained by fitting the experimental measurement of the SQUID resolution, with $\alpha=2.5\times 10^{-5}$ a large-distance cutoff, and $F_0$ a normalisation constant. It is worth noting that the current variations on the lattice scale are much larger than the average current. We also compared the results using a Gaussian filter, obtaining qualitatively similar behaviour.

\subsection*{Absence of Hall quantization for a clean system, and role of disorder in stabilising conductance plateau} In the next few paragraphs, we give an explicit account of how the genesis of the conductance plateau differs in basic ways between the QAHE and the QHE. This is a consequence  of the presence of counter-propagating states in the Chern band, which can spoil quantisation in the clean case, and need to be localised by disorder. This happens as follows. 
Consider a situation where the chemical potential $\mu$ lies below the bottom of the valence band 
near the edge, and above its top near the centre, of the sample. In the absence of disorder, and for an infinitely
long strip, translation symmetry along the $x$ direction implies that the momentum $k_x$ is a good quantum number.
The presence of a transverse confining potential along $y$ yields a discrete spectrum labelled with an integer quantum number $n$.
Let us denote by $\epsilon_{n}(k_x)$ the corresponding energy spectrum. In such quasi 1D geometry, the equation
$\epsilon_{n}(k_x)=\mu$ has a finite number of solutions $(n_i,k_{n,i})$, which each define  a particular conducting channel.
Recall that we have two spatial intervals either side of the centre of the strip where the chemical potential lies within the valence band. Assuming that quantum tunneling processes between these two intervals at energy $\mu$ can be neglected, we get half of the conducting channels with spatial support lying in the upper $y$ interval and the other half in the lower interval. Let us now concentrate on the former subset of channels. Their corresponding wave-functions are subject to a confining force directed towards the center of the strip.
The simultaneous presence of this transverse force and of a non-zero Chern number, $\mathrm{Ch}$, in the valence band creates 
a left-right asymmetry among these channels. Denoting by $N$ (resp. $M$) the number of channels with a positive (resp. negative) group velocity $d\epsilon_{n}(k_x)/d k_x$, we have the general relation $N-M= \pm \mathrm{Ch}$, where the sign
depends on the direction of the transverse force along the $y$ axis. This relation follows from the existence
of a spectral flow $\epsilon_{n}(k_x+2\pi/a)=\epsilon_{n \pm \mathrm{Ch}}(k_x)$.

In order to evaluate the four-terminal Hall conductance of a finite clean strip, we need to specify the four terminal transmission
matrix~\cite{Buttiker_edge}. Let us label the four reservoirs by letters as indicated on Fig.~\ref{fig:dichotom}. To simplify this discussion, we assume that these transmission matrix elements are
arranged in four groups:
$T_{\mathrm{CD}}\simeq T_{\mathrm{DC}} = (N+M)(1-\epsilon)$ (direct transmission),
$T_{\mathrm{AC}}\simeq T_{\mathrm{DA}}\simeq T_{\mathrm{BD}}\simeq T_{\mathrm{CB}} = N \epsilon$ (chiral propagation),
$T_{\mathrm{BC}}\simeq T_{\mathrm{DB}}\simeq T_{\mathrm{AD}}\simeq T_{\mathrm{CA}} = M \epsilon$ (chiral counter-propagation),
$T_{\mathrm{AB}}\simeq T_{\mathrm{BA}}\simeq 0$ (absence of direct scattering between lateral voltage probes for a large enough sample).
The small parameter $\epsilon$ indicates the probability that a wave packet injected in any of the $2N$ propagating channels
or any of the $2M$ counter-propagating ones will tunnel into one of the two lateral voltage probes. Because here we focus
on an ideal system, we neglect in this model the contact resistances at the left (C) and right (D) reservoirs.
This implies that all incoming wave packets from these reservoirs are eventually transmitted to other reservoirs,
and this translates into $\sum_{\mathrm{R'} \neq \mathrm{R}} T_{\mathrm{R'R}} = M+N$ for $\mathrm{R}=\mathrm{C}, \mathrm{D}$. Following standard procedure, we obtain the Hall resistance, in units of~$h/e^2$,
\begin{equation}
    R_{\mathrm{H}}=\frac{N-M}{(N+M)^2 - 2MN \epsilon}.
\end{equation}
This formula is quite instructive. First, it shows that no quantization is expected in the presence of counter-propagating channels ($M N \neq 0$), because the Hall resistance depends on the value of the tunnel probability~$\epsilon$. The quantized value $R_{\mathrm{H}}=1/N$,
independent of $\epsilon$, is recovered only in the absence of
counter-propagating channels. The ubiquity of such channels for clean topological insulators in the presence of
a smooth confining potential is in sharp contrast with the situation for the QHE Landau levels in a strong magnetic field, where
counter-propagating edge channels are entirely absent on this level of description for a smooth confining potential.
\begin{figure}
\centering
\includegraphics[width=\columnwidth]{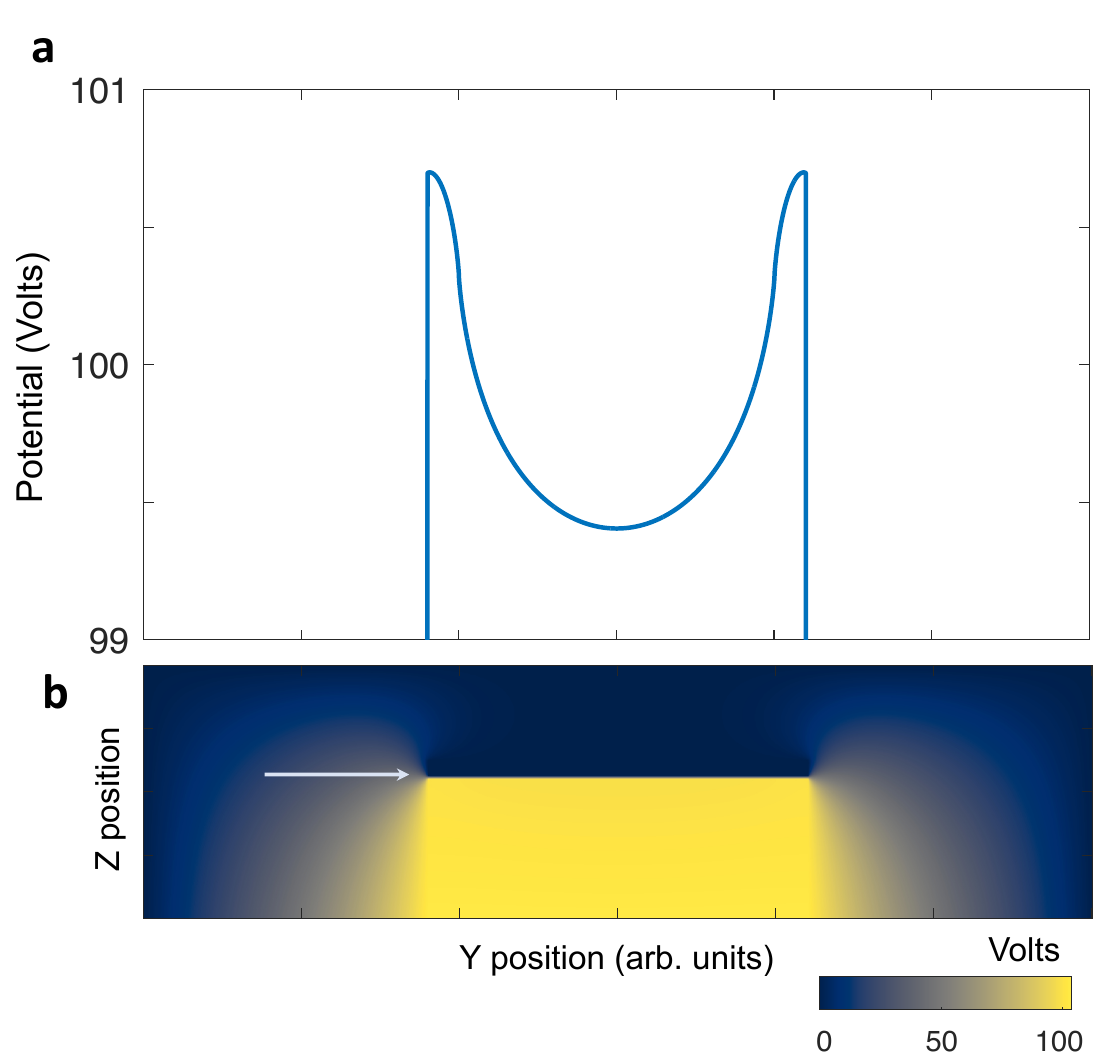}
\caption{\textbf{Origin of a quadratic electrostatic confining potential:} electrostatic potential calculated for the device geometry related to the experiment, Ref.~\cite{cornell_CIcurrent}. \textbf{(a)}: potential profile at the position of Chern insulator, which is separated by a thin layer of Aluminium oxide from the top metallic gate, and positioned in the proximity of a strontium titanate substrate. This is indicated by the arrow in panel \textbf{(b)}, which displays a cut of the three-dimensional sample geometry transverse to its long ($x$) direction. We set the voltage of the bottom gate to $110\,V$; the dielectric constants of aluminium oxide/strontium titanate were taken as $10$ and $10^4$ respectively, and the ratio of the width of the top gate/substrate to the width of the Chern insulator $~1.2$.  Note that for this backgate voltage, we obtain a potential change at the sample location of the order of $1\,V$.}
    \label{fig:parabola}
\end{figure}

Therefore, the role of disorder in the stabilization of sharply quantized Hall plateaus is
significantly different in these two situations. For the QAHE, we have just seen that disorder
is necessary to eliminate counter-propagating channels via Anderson localisation,
and thus to ensure a quantized Hall response carried by meandering channels. In the QHE, the clean system exhibits perfectly
quantized Hall conductance as long as the chemical potential lies between two consecutive Landau levels.
For a fixed electronic density, this occurs only for a discrete set of magnetic field values, corresponding to an integer number of filled Landau levels. Therefore the quantized Hall plateaus have a vanishing width, as a function of external magnetic field for a clean system.
The main role of disorder in the QHE, as has been realised for a long time, is to generate plateaux of a finite width, by allowing the chemical potential to change smoothly inside the gaps as the magnetic field is varied.

\subsection*{Origin of the confinement potential} The confinement potential, in particular its interplay with the finite bandwidth of the Chern insulator, plays an important role for the current distribution. In order to show how a parabolic potential can appear in an electrostatic problem, and to estimate its size, we have calculated the electrostatic potential in a model of the experimental system using a finite element method. Appendix Fig.~\ref{fig:parabola} displays the results of such a calculation, which yields a quadratic confining potential of the form utilised in our modelling. The difference between its maximum and minimum turns out to be of the order $1\,eV$ for a backgate voltage of $100\,V$.  We note that, depending on the device configuration, one can obtain both signs of the parabola.

\begin{figure}
\centering
\includegraphics[width=\columnwidth]{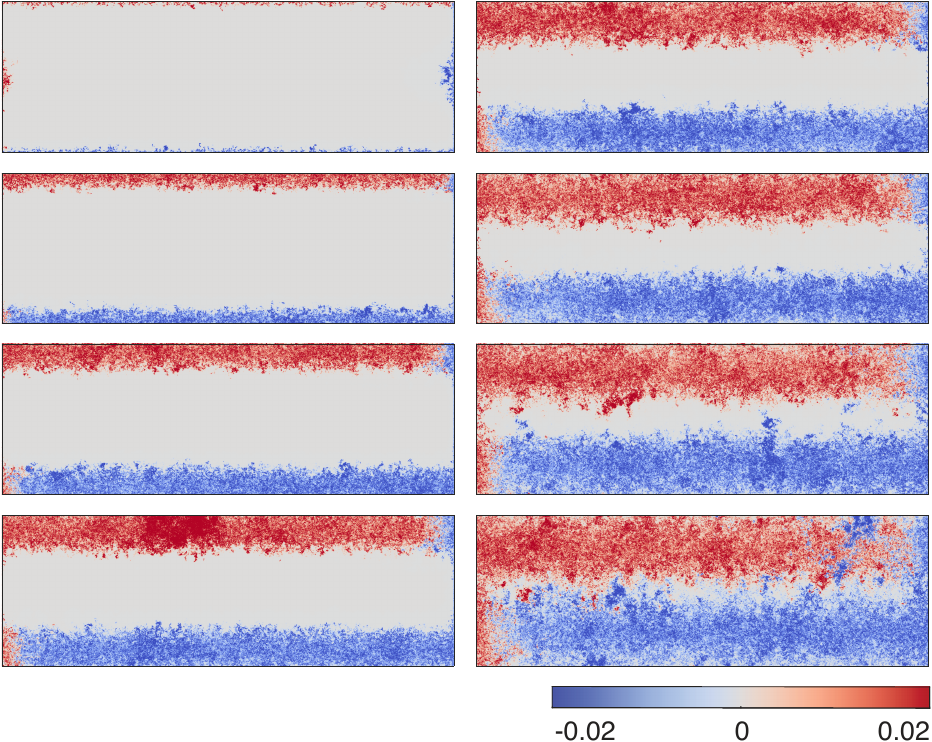}
\caption{\textbf{Non-equilibrium particle density} (excess density) generated by the application of a chemical potential difference to the left/right lead of the system corresponding to and for the same parameters as the panels in Fig.~2. This shows the spatial distribution of particles/holes injected from the leads. At large chemical potentials, only the states within narrow regions of the edge carry excess density. These compressible regions become progressively wider as we decrease the chemical potential, and merge when the latter is tuned to close to the top of the valence band.}
	\label{fig:density_panels}
\end{figure}

\subsection*{Phenomenological model for the effect of interactions on the density and current distributions}
Our above interpretation of current profiles in Ref.~\cite{cornell_CIcurrent} is based on a simple phenomenological model. 
Without Coulomb interactions, the quantized non-equilibrium current is carried by additional electrons and holes in the respective compressible strips hosting the meandering channels, as shown in Fig.~\ref{fig:density_panels}. With Coulomb interactions, this pair of oppositely charged strips creates an electrostatic
field across the central incompressible region, leading to an additional drop $\delta V_{\mathrm{I}}$ in electrochemical potential between the two strips.
A total current  $I_{\mathrm{I}}$ then also flows in this central incompressible region, characterized by the same quantized Hall conductance as the meandering channels.

Based on this picture, we introduce an effective model, in which we assume translation invariance along the wire (obtained for example after averaging over disorder). 
Let us denote by $\phi(y)$ the local electrostatic potential seen by electrons, the sum of the equilibrium
potential $\phi_{0}(y)$ and the induced potential $\phi_{\mathrm{ind}}(y)$ created by the charged compressible strips that appear under finite dc voltage bias. We assume that the particle density  at energy $E$ is a function of the kinetic energy in the valence band, which is given by $E+e\phi(y)$. Specifically, 
$n(y,E)=f(E+e\phi(y))/a_x a_y$, where $f$ is a positive function with unit total weight, and support lying in the 
$[0,W]$ interval,  $W$ being the energy width of the valence band, and $a_x$ and $a_y$ are the lattice constants along $x$ and $y$ directions. A second key assumption is that the local current density
$j(y,E)$ carried by a state at energy $E$ is given by:
$j(y,E)=(e^{2}/h) f(E+e\phi(y)) \phi'(y)$. In   equilibrium,
local particle, $n(y)$, and current density, $j(y)$, are obtained
by integrating the energy-resolved functions $n(y,E)$ and $j(y,E)$ over
$E$ up to the equilibrium chemical potential $\mu$. The incompressible regions
correspond to $\mu+e\phi_0(y) > W$, so the corresponding local density
$n_0(y)=1/a_x a_y$ as expected for a filled valence band. 

\begin{figure}
\includegraphics[width=\columnwidth]{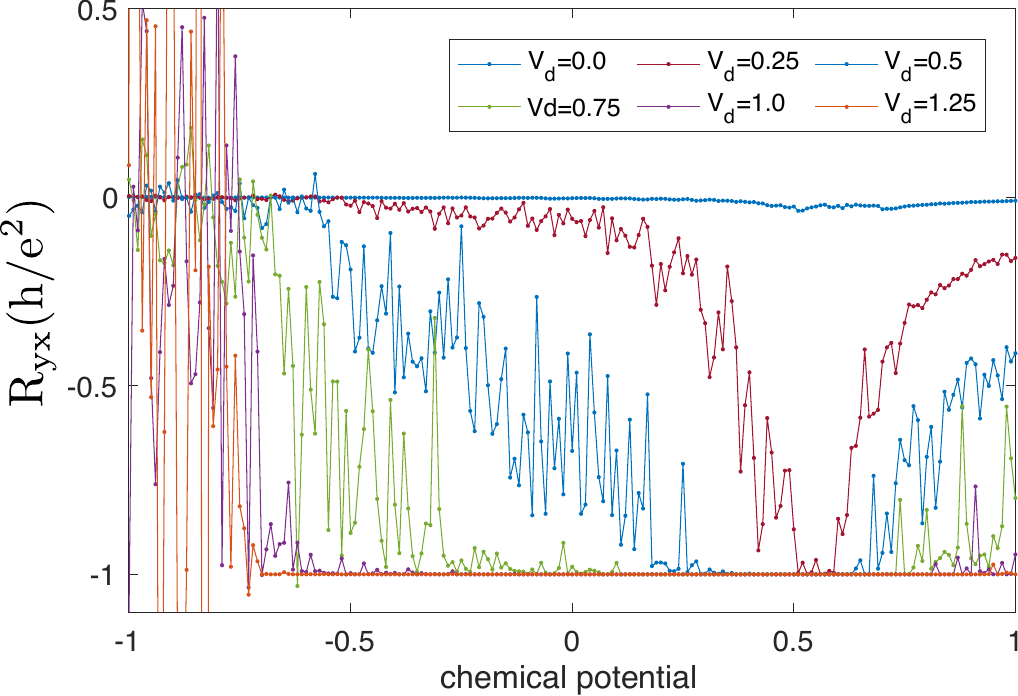}
\caption{\textbf{Conductance quantisation in presence of disorder:}
Disorder is needed to yield  conductance quantisation over a wide range of chemical potentials. The transverse conductance $R_{yx}$ was obtained by attaching two additional lateral leads to top and bottom of the system, e.g. in a four-terminal geometry in presence of a transverse parabolic potential with strength $V_p=5\times 10^{-6}$. The conductance was calculated using KWANT following theoretical approach of Ref.~\cite{Buttiker_edge}. We used a single disorder realisation in the calculations for each curve (no disorder averaging), which is the reason for the large oscillations. The conductance develops a quantized plateau, which widens with increasing disorder strength, and at large values of disorder the width of the quantized plateau is of the order of the band gap in absence of disorder.}
	\label{fig:quantisation}
\end{figure}

In the presence of a finite voltage bias, we integrate energies up to
$\mu_{b}=\mu+\delta \mu$ for $y$ corresponding to the bottom half of the strip, and
to $\mu_{t}=\mu-\delta \mu$ for $y$ in the top half. This model gives a quantized
Hall response for the total current $I$, obtained by integrating $j(y)$ over the whole width
of the wire, as $I=(e/h)(\mu_{b}-\mu_{t})$, or equivalently 
$I=(e^2/h)(\phi_{t}-\phi_{b})$ in terms of the total electrochemical potential difference across the wire. We  denote by $\delta n(y)$ and $\delta j(y)$ the change in local
particle and current densities after introducing the bias $\delta \mu$. 

The induced electrostatic potential $\phi_{\mathrm{ind}}(y)$ is 
\begin{equation}
   \phi_{\mathrm{ind}}(y) = - \frac{e}{2 \pi \epsilon_0} \int dy'\,\delta n(y') \log
   \frac{\sqrt{(y-y')^2+d^2}}{|y-y'|}.
   \label{eqphi_ind}
\end{equation} 
where we assume an infinitely long wire, and the presence of a perfectly conducting top electrode located at distance $d$ above the wire. In the Cornell experiment $d=40$nm.
To first order in $\delta \mu$, 
\begin{equation}
 \delta n(y)= \frac{f(\mu+e\phi_0(y))}{a_x a_y}[(\theta(-y)-\theta(y))\delta \mu + e\phi_{\mathrm{ind}}(y)].
 \label{eqdelta_n}
\end{equation}
To solve the coupled equations (\ref{eqphi_ind}) and (\ref{eqdelta_n}), one may eliminate either
$\delta n(y)$ or $\phi_{\mathrm{ind}}(y)$ from one of these equations, yielding an inhomogeneous
linear integral equation for the remaining variable, which we have solved numerically.
After this step, the non equilibrium current in the incompressible region is 
\begin{equation}
\delta j(y)=\frac{e^2}{h}\phi'_{\mathrm{ind}}(y).
\end{equation}
while in the lateral compressible strips
\begin{multline}
\delta j(y) = \frac{e^2}{h}f(\mu+e\phi_0(y))[(\theta(-y)-\theta(y))\delta \mu + e\phi_{\mathrm{ind}}(y)]\phi'_{0}(y) \\ + \frac{e^2}{h}\phi'_{\mathrm{ind}}(y) \int_{0}^{\mu+e\phi_0(y)}f(E) dE.
\end{multline}
The physical interpretation of this expression is quite simple. The first term arises from the change in the
local electronic density in the non-equilibrium state as compared to the equilibrium situation, in the presence of
the {\it un}perturbed confining potential and hence unperturbed local anomalous velocity field. The second term may be regarded as the response (in the spirit of the Kubo formula) of a locally partially filled (compressible strips)
or completely filled (incompressible strips) valence band to the additional electrostatic potential created
by the change in the charge distribution.

\begin{figure}
\centering
\includegraphics[width=\columnwidth]{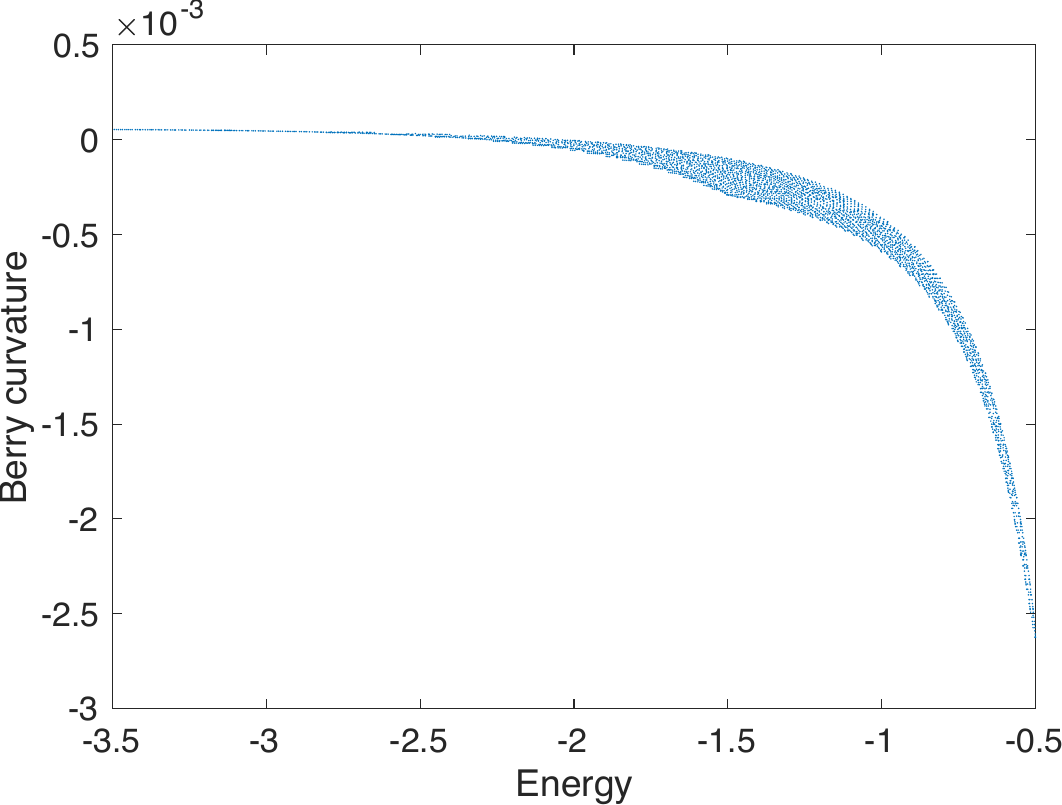}%
	\caption{\textbf{Energy dependence of Berry curvature} for the model of a clean Chern insulator in the valence band: note the sign change, implying a preponderance of opposite Berry curvatures on opposite sides of the conduction channel. This figure was obtained by a scatter plot of the curvature vs energy taken at all values of momentum in the Brillouin zone (system size $100\times300$ with periodic boundary conditions in both directions).}
	\label{fig:berry_curvature}
\end{figure}

To estimate the strength of the Coulomb interaction, we  first neglect
$\phi_{\mathrm{ind}}(y)$ in Eq.~(\ref{eqdelta_n}), and compute the maximal absolute value $\kappa(\mu)$ of the
ratio $e\phi_{\mathrm{ind}}(y)/\delta \mu$. It is reached when $y$ lies near the centers of the compressible
strips. Assuming that they are much larger than the top gate distance $d$, we estimate this
ratio to be of the order of $4 (R_y/W)(a_0 d/a_x a_y)$, where $R_y$ is the Rydberg energy and $a_0$
is the Bohr radius. We get a rather large number (on the order of $10^3$) for the Cornell experiment.

Coulomb interactions modify the current profile.
Most notable is a reduction of the non-equilibrium charge density $\delta n(y)$
for a given $\delta \mu$ because the Coulomb repulsion among particles on a given
compressible strip increases their total potential energy. Assuming for simplicity that only the
overall scale of $\delta n(y)$ is modified, but not its shape, this can be captured by a renormalization of the bias $\delta \mu$ into a much smaller $\delta \mu^*$, defined
by $\delta \mu^*=\delta \mu - \kappa(\mu) \delta \mu^*$, so $\delta \mu^*=\delta\mu /(1+\kappa(\mu))$. The new
maximal value of $e\phi_{\mathrm{ind}}(y)/\delta \mu$ is now equal to 
$\kappa(\mu)_{\mathrm{eff}}=\kappa(\mu)/(1+\kappa(\mu))$, which is
very close to, but smaller than 1, in the physically relevant case where $\kappa(\mu)$ is large.
Note that $\kappa(\mu)$ is linear in the strength $R_y/W$ of the Coulomb interaction.
So $\delta \mu^*$ decreases as $W/R_y$ when interactions are large, but the 
$e\phi_{\mathrm{ind}}(y)/\delta \mu$ ratio scales to a finite limit value at large
$R_y/W$. This value depends mostly on geometric parameters such as the widths of the
compressible strips, their separation, and the effective range $d$ of the screened
Coulomb potential.

The reduction of the non-equilibrium charge density is not uniform across the compressible strips, so that $\delta n(y)$ becomes more uniform, and hence effectively wider, than the 
non-interacting envelope function $f(\mu+e\phi_0(y))$. These effects are clearly seen in the numerical results shown in Fig.~\ref{fig:currentpatternCoulomb}.

\section*{Acknowledgements}

We thank John Chalker, Matt Ferguson, Curt von Keyserlingk,  Klaus von Klitzing, Katja Nowack and Boris Shklovskii for discussions, and F.~Evers for comments on the manuscript. We are grateful to Bennet Becker and Hubert Scherrer-Paulus at the IT department of MPIPKS for their help with
parallelisation of our computer simulations. B.D.~and D.K.~thank MPIPKS for its generous hospitality during several extended visits, which were crucial for the realization of this project. This work was in part supported by the Deutsche Forschungsgemeinschaft under grant  cluster of excellence ct.qmat (EXC 2147, project-id 390858490). D.K.~acknowledges support from Labex MME-DII grant ANR11-LBX-0023, and funding under The Paris Seine Initiative Emergence programme 2019.\\

\bibliography{references}

\end{document}